\documentclass[aps,reprint,twocolumn,superscriptaddress,citeautoscript,showpacs,amsmath,amssymb,prl]{revtex4-1}
\usepackage{graphicx}
\usepackage{dcolumn}
\usepackage{bm}
\usepackage{times}
\usepackage{color}
\begin{document}
\title{Fragile antiferromagnetism in the heavy-fermion compound YbBiPt}

\author{B. G. Ueland}
\affiliation{Ames Laboratory, U.S. DOE, Iowa State University, Ames, Iowa 50011, USA}
\affiliation{Department of Physics and Astronomy, Iowa State University, Ames, Iowa 50011, USA}

\author{A. Kreyssig}
\affiliation{Ames Laboratory, U.S. DOE, Iowa State University, Ames, Iowa 50011, USA}
\affiliation{Department of Physics and Astronomy, Iowa State University, Ames, Iowa 50011, USA}

\author{K. Proke\v{s}}
\affiliation{Helmholtz-Zentrum Berlin f{\"{u}}r Materialien und Energie, Hahn-Meitner-Platz 1, 14109 Berlin, Germany}

\author{J. W. Lynn}
\affiliation{NIST Center for Neutron Research, National Institute of Standards and Technology, Gaithersburg, Maryland 20899, USA}

\author{L. W. Harriger}
\affiliation{NIST Center for Neutron Research, National Institute of Standards and Technology, Gaithersburg, Maryland 20899, USA}

\author{D. K. Pratt}
\affiliation{NIST Center for Neutron Research, National Institute of Standards and Technology, Gaithersburg, Maryland 20899, USA}

\author{D. K. Singh}
\affiliation{NIST Center for Neutron Research, National Institute of Standards and Technology, Gaithersburg, Maryland 20899, USA}
\affiliation{Department of Materials Science and Engineering, University of Maryland, College Park, MD 20742, USA}

\author{T. W. Heitmann}
\affiliation{The Missouri Research Reactor, University of Missouri, Columbia, Missouri 65211, USA}

\author{S. Sauerbrei}
\affiliation{Ames Laboratory, U.S. DOE, Iowa State University, Ames, Iowa 50011, USA}
\affiliation{Department of Physics and Astronomy, Iowa State University, Ames, Iowa 50011, USA}

\author{S. M. Saunders}
\affiliation{Ames Laboratory, U.S. DOE, Iowa State University, Ames, Iowa 50011, USA}
\affiliation{Department of Physics and Astronomy, Iowa State University, Ames, Iowa 50011, USA}

\author{E. D. Mun}
\affiliation{Ames Laboratory, U.S. DOE, Iowa State University, Ames, Iowa 50011, USA}
\affiliation{Department of Physics and Astronomy, Iowa State University, Ames, Iowa 50011, USA}

\author{S. L. Bud'ko}
\affiliation{Ames Laboratory, U.S. DOE, Iowa State University, Ames, Iowa 50011, USA}
\affiliation{Department of Physics and Astronomy, Iowa State University, Ames, Iowa 50011, USA}

\author{R. J. McQueeney}
\affiliation{Ames Laboratory, U.S. DOE, Iowa State University, Ames, Iowa 50011, USA}
\affiliation{Department of Physics and Astronomy, Iowa State University, Ames, Iowa 50011, USA}

\author{P. C. Canfield}
\affiliation{Ames Laboratory, U.S. DOE, Iowa State University, Ames, Iowa 50011, USA}
\affiliation{Department of Physics and Astronomy, Iowa State University, Ames, Iowa 50011, USA}

\author{A. I. Goldman}
\affiliation{Ames Laboratory, U.S. DOE, Iowa State University, Ames, Iowa 50011, USA}
\affiliation{Department of Physics and Astronomy, Iowa State University, Ames, Iowa 50011, USA}

\date{\today}

\begin{abstract}
We report results from neutron scattering experiments on single crystals of YbBiPt that demonstrate antiferromagnetic order characterized by a propagation vector, $\bm\tau_{\rm{AFM}}$ = ($\frac{1}{2} \frac{1}{2} \frac{1}{2}$), and ordered moments that align along the [1 1 1] direction of the cubic unit cell.  We describe the scattering in terms of a two-Gaussian peak fit, which consists of a narrower component that appears below $T_{\rm{N}}~\approx~0.4$ K and corresponds to a magnetic correlation length of $\xi_{\rm{n}} \approx$ 80~$\rm{\AA}$, and a broad component that persists up to $T^*\approx$ 0.7~K and corresponds to antiferromagnetic correlations extending over $\xi_{\rm{b}} \approx$ 20~$\rm{\AA}$.  Our results illustrate the fragile magnetic order present in YbBiPt and provide a path forward for microscopic investigations of the ground states and fluctuations associated with the purported quantum critical point in this heavy-fermion compound.
\end{abstract}

\maketitle

Unusual magnetic behavior may occur in heavy-fermion systems \cite{Stewart_1984} in close proximity to a magnetic quantum critical point (QCP) \cite{Stewart_2001,Stockert_2011,Si_2010_1} due to the entanglement of conduction electrons and localized moments and the competition between potential ground states \cite{Senthil_2004,Si_2010}.  Quantum phase transitions occur at $T$ = 0~K and are driven by some non-thermal parameter such as magnetic field or pressure \cite{Sachdev_1999}.  In strongly-correlated electron systems such as heavy-fermions compounds, quantum phase transitions may be accompanied by large changes in the Fermi-surface and can lead to non-Fermi liquid behavior, enhanced quantum fluctuations, and may result in superconductivity or other novel ground states \cite{Si_2010,Coleman_2007}.

Two scenarios are often discussed in the context of heavy-fermions with QCPs \cite{Stockert_2011}: (1) the conventional spin-density-wave (SDW) scenario where the quasiparticles are formed below the Kondo temperature ($T_{\rm{K}}$) and survive in the vicinity of the QCP yielding critical fluctuations localized at small regions of the Fermi-surface  \cite{Hertz_1976,Millis_1993}; and (2) the Kondo breakdown scenario \cite{Si_2010} where localization of the \emph{f} electrons at the QCP breaks the Kondo coupling yielding large changes of the Fermi-surface accompanied by a magnetic transition.  CeCu$_{2}$Si$_{2}$ \cite{Stockert_2004} and Ce$_{1-x}$La$_{x}$Ru$_{2}$Si$_{2}$ \cite{Knafo_2009} are cited as examples of materials described by the conventional SDW scheme, whereas CeCu$_{6-x}$Au$_x$ \cite{Si_2001} and YbRh$_{2}$Si$_{2}$ \cite{Friedemann_2009} provide examples relevant to the latter scenario.

Experiments on YbRh$_{2-x}$Ir$_{x}$Si$_{2}$ have shown that substituting 6\% Ir for Rh detaches the Kondo-breakdown point from the QCP resulting in an extended intermediate-field range of non-Fermi liquid (NFL) behavior, characteristic of a "spin-liquid"-type ground state \cite{Friedemann_2009}. The stoichiometric compound YbAgGe \cite{Budko_2004} also exhibits an extended region of NFL behavior with applied magnetic field \cite{Niklowitz_2006,Schmiedeshoff_2011}.  However, accessing the ordered magnetic state close to the QCP and studying the evolution of the microscopic magnetic correlations in the NFL regime is complicated by either the requirement of attaining extremely low temperatures ($T_{\rm{N}}~\approx~0.05$ K) for YbRh$_{2}$Si$_{2}$ \cite{Ishida_2003,Stock_2012}, or by a complex series of magnetic transitions with applied field  for YbAgGe \cite{Schmiedeshoff_2011}.  YbBiPt offers an important alternative stoichiometric system with: (1) the simplicity of a cubic lattice; (2) temperatures and fields that are low, but readily achievable for scattering measurements close to the QCP; and (3) a rather simple $H-T$ phase diagram with an extended region of NFL behavior \cite{Canfield_1991,Fisk_1991,Movshovich_1994,Canfield_1994,Mun_2013,Robinson_1994}.

YbBiPt belongs to the series of cubic half-Heusler (space group $F\overline{4}3m$) $R$BiPt compounds (\emph{R} = rare earth)  \cite{Robinson_1994,Wosnitza_2006,Kreyssig_2011}, with the magnetic Yb ions forming a face-centered-cubic magnetic sublattice. Its discovery generated strong interest due to its extraordinary Sommerfield coefficient ($\gamma~\approx$ 8~J/mol-K$^2$) and classification as a heavy-fermion compound \cite{Canfield_1991,Fisk_1991,Movshovich_1994,Canfield_1994}. All of its relevant energy scales including the Kondo temperature ($T_{\rm{K}} \approx$ 1~K) that describes the magnetic coupling between the localized and itinerant moments, the Weiss temperature ($\theta_{\rm{W}} \approx$~-2~K) that describes the mean-field magnetic exchange strength, the N\'{e}el temperature for the proposed spin-density-wave order ($T_{\rm{N}}$ = 0.4~K), and the crystalline electric field splitting ($\Delta E <$ 1~meV) are small and comparable, suggesting a complex interplay of competing interactions at low temperature. It has also been suggested that YbBiPt offers the realization of a topological Kondo insulator \cite{Chadov_2010}.

Much of the recent attention on YbBiPt has focused on the possibility of a magnetic-field-tuned antiferromagnetic (AFM) QCP occurring at a low critical magnetic field of $\mu_{0}H_{\rm{c}}$ = 0.4~T \cite{Mun_2013}. Thermodynamic and transport measurements in ambient field suggest that YbBiPt manifests AFM order below $T_{\rm{N}}$ = 0.4~K. In particular, a clear anomaly is observed at $T_{\rm{N}}$ in electrical resistance data that is consistent with spin-density-wave type AFM order that partially gaps the Fermi surface \cite{Canfield_1991}. This feature is strongly suppressed upon the application of a modest magnetic field ($\mu_{0}H > \mu_{0}H_{\rm{c}}$), and non-Fermi liquid behavior is found for $\mu_{0}H_{\rm{c}} < H <$ 0.8~T, followed by Fermi-liquid behavior for $\mu_{0}H >$ 0.8~T \cite{Mun_2013}. Although the locations of the field-induced phase transitions and Fermi-liquid behavior have been mapped out \cite{Mun_2013}, it is not yet clear whether YbBiPt is best described by the conventional SDW or the Kondo breakdown scenario.  

It also is notable that scattering measurements over the past 22 years have failed to identify magnetic ordering in powder \cite{Robinson_1994} or single crystal samples, leading to uncertainty regarding the true nature of the proposed AFM transition that is somewhat reminiscent of the "hidden order" paradox in URu$_2$Si$_2$ \cite{Mydosh_2011}.  Furthermore, muon spin-relaxation ($\mu$SR) measurements have found evidence of spatially inhomogeneous and disordered magnetism in powder samples \cite{Amato_1992} which suggests that any magnetic order in YbBiPt is likely quite fragile. Clarifying the nature of the transition at $T$ = 0.4~K in YbBiPt represents a key step towards performing microscopic investigations of magnetism close to the QCP.

Here, we present results from neutron scattering experiments on single crystals of YbBiPt that identify and characterize the low temperature AFM order by the magnetic propagation vector $\bm\tau_{\rm{AFM}}$ = ($\frac{1}{2}~\frac{1}{2}~\frac{1}{2}$) and moments collinear with $\bm\tau_{\rm{AFM}}$.  We further show that the observed magnetic scattering can be modeled by a two-Gaussian peak fit consisting of a narrower Gaussian component that appears below $T_{\rm{N}}$ with a magnetic correlation length of $\xi_{\rm{n}} \approx$ 80~$\rm{\AA}$, and a broader Gaussian component that persists up to $T^* \approx$ 0.7~K that is consistent with short-range AFM correlations occurring over $\xi_{\rm{b}} \approx$ 20~$\rm{\AA}$.  We suggest that the narrower and broad components of the scattering illustrate the competition among the low-energy magnetic interactions and lend themselves to a picture of fragile magnetic order occurring at low temperature.

\begin{figure}
\centering
\includegraphics[width=1\linewidth]{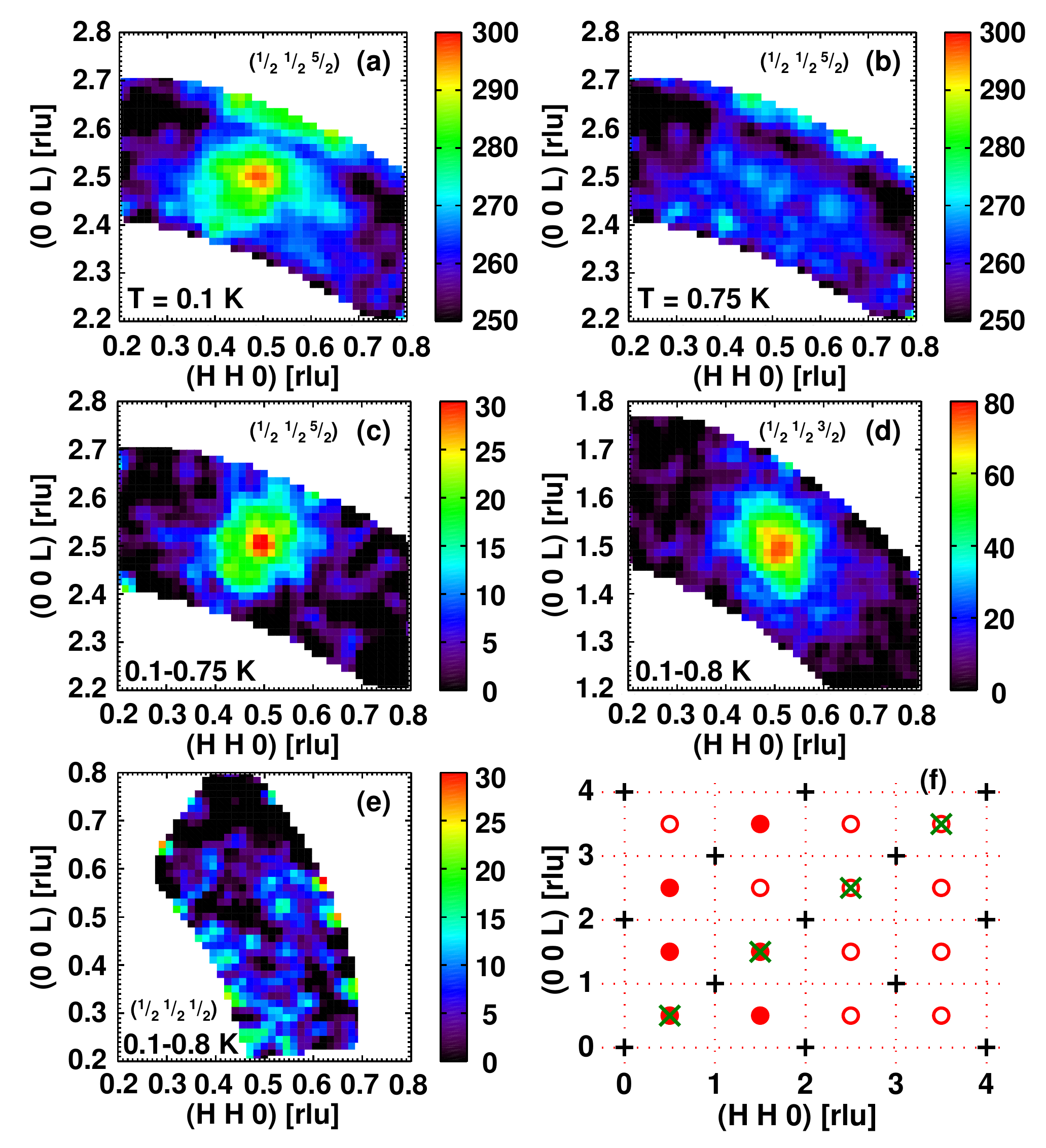}
\caption{(color online) Contour plots of diffraction data taken for points in the ($H H L$) plane corresponding to the antiferromagnetic propagation vector $\bm\tau_{\rm{AFM}}$ = ($\frac{1}{2} \frac{1}{2} \frac{1}{2}$).  The intensity of the scattering is indicated by color. (a) Data for the ($\frac{1}{2} \frac{1}{2} \frac{5}{2}$) position for $T$ = 0.1~K and (b) $T$ = 0.75~K, and (c) after subtracting $T$ = 0.1~K data by the $T$ = 0.75~K data.  Panels (d) and (e) show data for the  ($\frac{1}{2} \frac{1}{2} \frac{3}{2}$) and ($\frac{1}{2} \frac{1}{2} \frac{1}{2}$) positions, respectively, after subtracting the $T$ = 0.1~K data by the corresponding $T$ = 0.8~K data.  (f) Diagram of the ($H H L$) reciprocal lattice plane for YbBiPt.  Nuclear Bragg points are indicated by black crosses, and possible magnetic Bragg points are indicated by circles.  Solid circles correspond to measured points, and points where the intensity of the magnetic scattering is zero are marked with $\times$'s.  Dashed lines indicate the magnetic Brillouin zones.}
\label{Fig1}
\end{figure}

\begin{figure}[t!]
\centering
\includegraphics[width=1\linewidth]{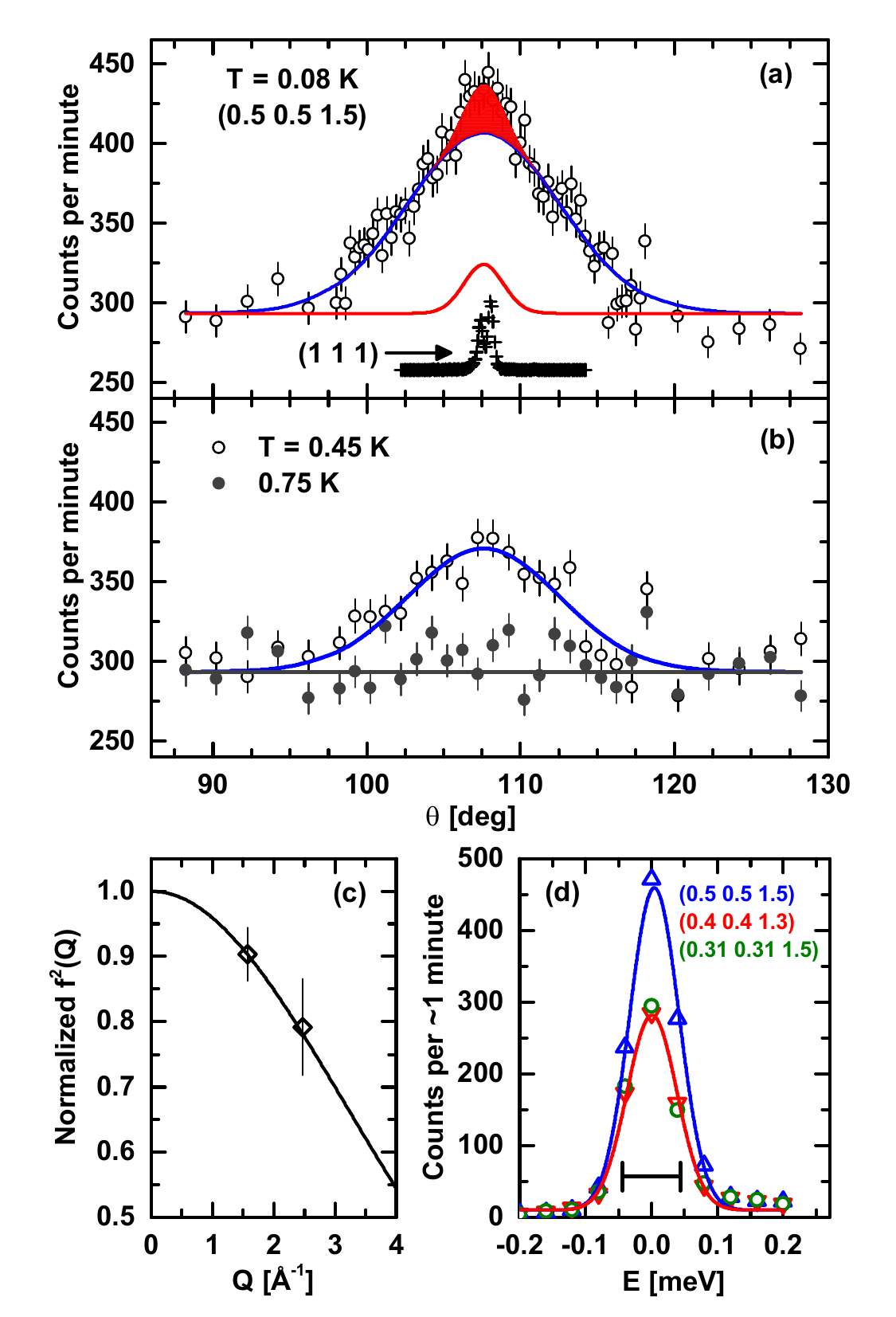}
\caption{(color online) Detailed scattering data for ($\frac{1}{2} \frac{1}{2} \frac{3}{2}$). (a) Data from a rocking scan taken at $T$ = 0.08~K.  Blue and red lines show the broad and narrower components of the two-Gaussian peak fit, respectively, and the shaded red area corresponds to scattering contributed by the narrower Gaussian peak.  Black crosses at the bottom show scaled data from a rocking scan at the (1 1 1) nuclear Bragg peak, which is split due to a small misalignment of the two co-aligned crystals. (b) Data from rocking scans taken at $T$ = 0.45~K (open circles) and 0.75~K (filled dark gray circles).  The blue curve represents a fit to a single Gaussian peak, while the dark gray line depicts the background.  (c) $Q$-dependence of the scattering at ($\frac{1}{2} \frac{1}{2} \frac{3}{2}$) and ($\frac{1}{2} \frac{1}{2} \frac{5}{2}$) compared to the square of the Yb$^{3+}$ magnetic form factor (solid line). (d) Energy dependence of the scattering from constant $\textbf{Q}$ scans at $T$ = 0.08~K for \textbf{Q} = ($\frac{1}{2} \frac{1}{2} \frac{3}{2}$) (upward pointing triangles), and background scans at two different \textbf{Q}-positions (circles and downward pointing triangles).  Lines represent fits to Gaussian peaks, and the energy resolution (FWHM) is indicated by the horizontal line.  The shaded area corresponds to elastic magnetic scattering at ($\frac{1}{2} \frac{1}{2} \frac{3}{2}$). Uncertainties represent one standard deviation.}
\label{Fig2}
\end{figure}

Single crystals of YbBiPt were grown out of a Bi flux as described previously \cite{Mun_2013} and ranged in mass from several hundred mg to nearly 2 g.  Several samples with total masses of 1 - 3~g and total mosaic spreads of $\approx 1^{\circ}$ FWHM were assembled for neutron scattering experiments using either one crystal or two co-aligned crystals.  Given the strong sensitivity of the samples to pressure and strain \cite{Movshovich_1994,Mun_2013}, several methods and glues [an amorphous fluoropolymer (CYTOP) or dental glue (HBM X60)] were used to fix the crystals to a Cu sample holder which was then thermally anchored to the bottom of a dilution refrigerator. For the samples attached with the fluoropolymer, Cu wire was loosely wrapped around the crystals and anchored to the sample holder to ensure mechanical stability.

Neutron scattering experiments were performed on the E-4 two-axis diffractometer at Helmholtz-Zentrum Berlin, and the SPINS cold-neutron and BT-7 thermal-neutron triple-axis spectrometers \cite{Lynn_2012} at the NIST Center for Neutron Research.  Incident neutrons with wavelengths of $\lambda$ = 2.451, 5.504, and 2.359~{\AA}, for E-4, SPINS, and BT-7, respectively,  were selected by a pyrolitic graphite (PG) monochromator, and PG or liquid nitrogen cooled Be filters were inserted to reduce contamination from higher-order wavelengths.  A 40$^{\prime}$ or  80$^{\prime}$ S\"{o}ller collimator was used between the monochromator and sample and a 120$^{\prime}$ (E-4 and SPINS) or 80$^{\prime}$  (BT-7) radial collimator was placed immediately after the sample.  BT-7 was operated in two-axis mode and both E-4 and BT-7 utilized position sensitive detectors.  On SPINS, a PG analyzer horizontally focused to a single $^3$He detector was used to select a fixed final neutron wavelength of $\lambda$ = 5.504~{\AA}.  The \textsc{LAMP} and \textsc{DAVE} software packages were used for data reduction \cite{LAMP,DAVE}.

Comprehensive searches for magnetic scattering in the ($H 0 L$) and ($H H L$) reciprocal lattice planes were undertaken on E-4 and resulted in the discovery of additional scattering below $T^*\approx$ 0.7~K at half-integer positions ($\frac{h}{2} \frac{h}{2} \frac{l}{2}$) with $h$ and $l$ odd integers and $h\ne l$.  Fig.~\ref{Fig1} shows diffraction data from rocking scans taken in the ($H H L$) plane using the BT-7 spectrometer.  Figure~\ref{Fig1}(a) shows a broad peak (in both the longitudinal and transverse directions) centered at ($\frac{1}{2} \frac{1}{2} \frac{5}{2}$) for $T$ = 0.1~K, and Fig.~\ref{Fig1}(b) shows that the peak is absent for $T$ = 0.75~K.  Figure~\ref{Fig1}(c) shows the same region after subtracting the $T$ = 0.75~K data from the  $T$ = 0.1~K data.  Figures~\ref{Fig1}(d) and (e) show similar plots of $T$ = 0.1~K data after subtracting $T$ = 0.8~K data for the ($\frac{1}{2} \frac{1}{2} \frac{3}{2}$) and ($\frac{1}{2} \frac{1}{2} \frac{1}{2}$) positions, respectively.  A broad peak centered at ($\frac{1}{2} \frac{1}{2} \frac{3}{2}$), similar to the one at ($\frac{1}{2} \frac{1}{2} \frac{5}{2}$), is observed in Fig.~\ref{Fig1}(d), whereas Fig.~\ref{Fig1}(e) shows that the peak is absent at the ($\frac{1}{2} \frac{1}{2} \frac{1}{2}$) position.  Although not shown, distinct but broad peaks similar to those in Figs.~\ref{Fig1}(c) and (d) were identified at the ($\pm\frac{1}{2}\pm\frac{1}{2}\mp\frac{3}{2}$) and ($\pm\frac{3}{2}\pm\frac{3}{2}\pm\frac{1}{2}$) positions, whereas no peaks were observed at the ($\pm\frac{3}{2}\pm\frac{3}{2}~\frac{3}{2}$) positions.  Figure~\ref{Fig1}(f) shows a diagram which summarizes our observations in the first quadrant of the ($H H L$) plane.   Since the intensity of the magnetic scattering is proportional to the component of the moment perpendicular to the  neutron momentum transfer, \textbf{Q}, the systematic absence of scattering at the ($\frac{1}{2} \frac{1}{2} \frac{1}{2}$) and ($\pm\frac{3}{2}\pm\frac{3}{2}~\frac{3}{2}$) positions indicates that the ordered moment is aligned along the [1~1~1] direction.   Hence, we conclude that the AFM propagation vector is $\bm\tau_{\rm{AFM}}$ = ($\frac{1}{2} \frac{1}{2} \frac{1}{2}$), and that the ordered moments are collinear with $\bm\tau_{\rm{AFM}}$.

To study the magnetic scattering in more detail, we performed measurements on a co-aligned sample on the SPINS spectrometer as a function of the neutron energy transfer $E$. Elastic  ($E$ = 0) data from rocking scans centered at the ($\frac{1}{2} \frac{1}{2} \frac{3}{2}$) position for $T$ = 0.08~K are shown in Fig.~\ref{Fig2}(a), and data for  $T$ = 0.45 and 0.75~K  are shown in Fig.~\ref{Fig2}(b).  For comparison, data from a rocking scan through the (1 1 1) nuclear Bragg peak are shown at the bottom of Fig.~\ref{Fig2}(a).  The magnetic scattering at $T$ = 0.08 and 0.45~K is much broader than the nuclear peak which indicates that a finite magnetic correlation length exists for the AFM order. For $T$ = 0.75~K the magnetic peaks are absent. Upon lowering the temperature below $T^*$, broad scattering appears that grows in intensity with decreasing temperature and is well described by a single Gaussian peak.  For temperatures below $T_{\rm{N}}~\approx$ 0.4~K, Fig.~\ref{Fig2}(a) shows that a single Gaussian peak no longer adequately describes the observed scattering since additional intensity with a narrower distribution is evident for $T$ = 0.08~K, and we describe the scattering data for $T < T_{\rm{N}}$ by a two-Gaussian peak fit that is the sum of a broad Gaussian component and a concentric narrower Gaussian component.  Since the centers and the FWHM's of the broad and narrower components of the two-Gaussian peak fit do not vary significantly with temperature, they were fixed to the values obtained at $T$ = 0.08~K [$\Delta\theta_{\rm{narrow}}$ = 3.2(9)$^\circ$ and $\Delta\theta_{\rm{broad}}$ = 12.5(9)$^\circ$]. A constant background determined at $T$ = 0.75~K has also been used.

Normalizing to the integrated intensity of the (1 1 1) nuclear reflection, and assuming equally populated magnetic domains and contributions from the full volume of the sample, we calculate the average magnetic moment at $T$ = 0.08~K associated with the total measured magnetic scattering at $\bm\tau_{\rm{AFM}}$ to be $\approx~0.8~\mu_{\rm{B}}$. At this point we do not attempt to partition the ordered moment between the broad and narrower components but note that the ratio of their integrated intensities is approximately 12:1. For the magnetic structure described above, we can also compare the $Q$-dependence of the scattering at the ($\frac{1}{2} \frac{1}{2} \frac{3}{2}$) and ($\frac{1}{2} \frac{1}{2} \frac{5}{2}$) magnetic Bragg positions with that expected for the Yb$^{3+}$ magnetic form factor [see Fig.~\ref{Fig2}(c)] and find good agreement. Taken together with the systematic absence of scattering at the ($\frac{1}{2} \frac{1}{2} \frac{1}{2}$) and ($\pm\frac{3}{2}\pm\frac{3}{2}~\frac{3}{2}$) positions, these data confirm the magnetic origin of the half-integer diffraction peaks.

\begin{figure}
\centering
\includegraphics[width=0.95\linewidth]{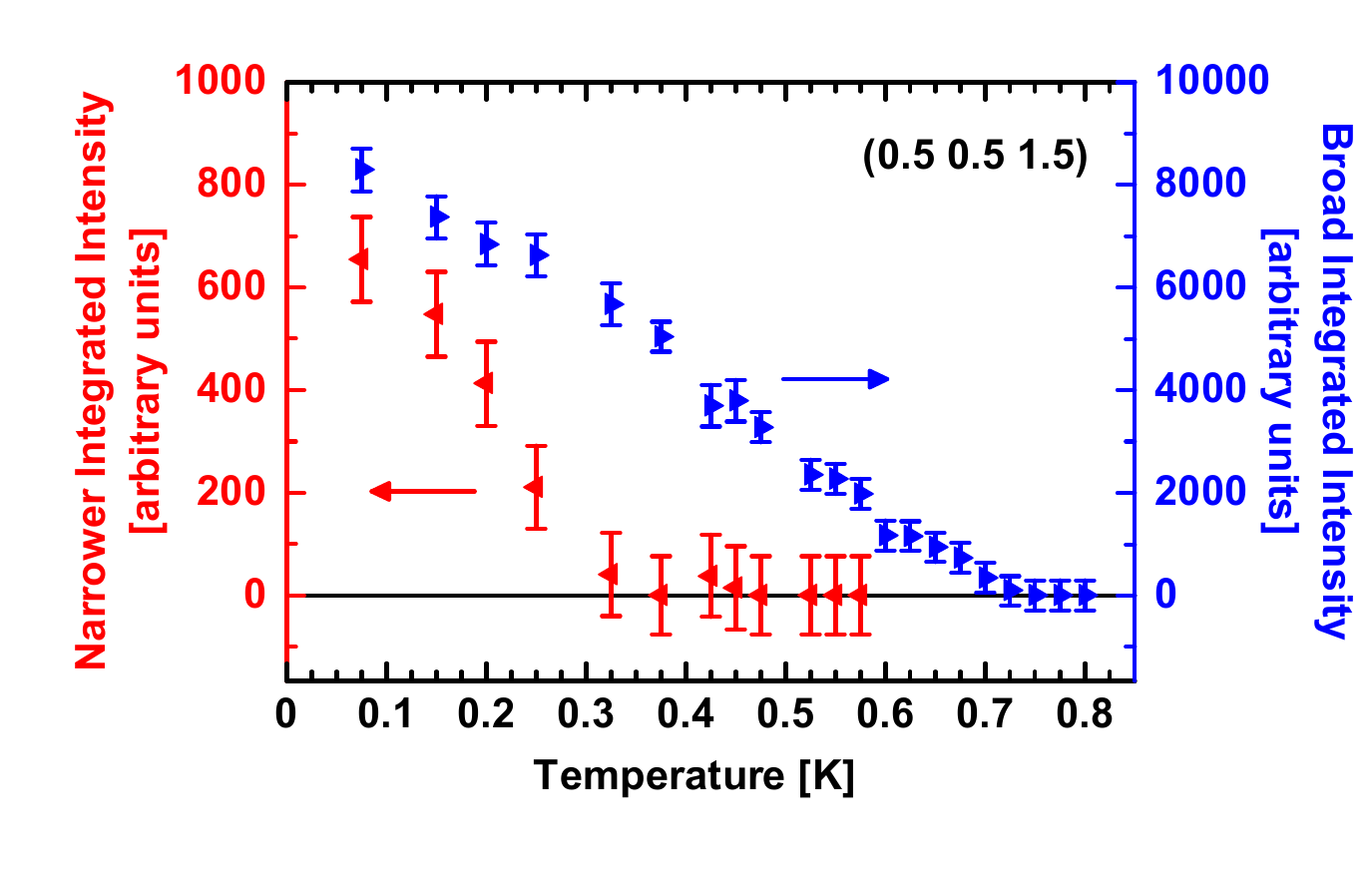}
\caption{(color online) Temperature dependence of the integrated intensities of the narrower (left axis) and broad (right axis) components of the two-Gaussian peak fits to the magnetic scattering at ($\frac{1}{2}~\frac{1}{2}~\frac{3}{2}$).}
\label{Fig3}
\end{figure}

We note that the data below $T_{\rm{N}}$ may also be described by a single Lorentzian-squared peak although the fit does not quite capture all of the low-temperature intensity at the center of the peak. Further measurements using significantly larger samples may be required to ultimately determine the most appropriate fitting function for the magnetic scattering. Nevertheless, the temperature dependence of the integrated intensities of the components of the two-Gaussian peak fit are shown in Fig.~\ref{Fig3} and suggest that the two-Gaussian peak fit captures the essential features of the scattering:  the narrower component decreases smoothly with increasing temperature and is absent above $T_{\rm{N}}$ $\approx$ 0.4~K, consistent with the bulk thermodynamic and transport measurement results, while the integrated intensity of the broad component also decreases smoothly with increasing temperature but persists up to $T^* \approx 0.7$~K.

The magnetic correlation lengths associated with the components of the two-Gaussian peak fit can be derived from the FWHM of the peaks in the rocking scan data and are $\xi_{\rm{n}} \approx 80$~{\AA} and $\xi_{\rm{b}} \approx 20$~{\AA} for the narrower and broad components, respectively. The presence of broad magnetic scattering and finite correlation lengths appears consistent with previous $\mu$SR measurements on powder samples which concluded that the ordered moment in YbBiPt is spatially inhomogeneous \cite{Amato_1992}. However, the $\mu$SR measurements were performed on powders raising the possibility of strain effects \cite{Mun_2013}.  The present measurements were performed on single crystals mounted to minimize or eliminate strain effects.

We believe that this unusual magnetic behavior is intrinsic to YbBiPt and does not arise from chemical or structural inhomogeneities because:  (1) The broad magnetic component has been found for all three sets of measured samples despite the crystals coming from different growth batches and despite different mounting methods.  (2) All samples measured present resolution limited nuclear diffraction peak widths in longitudinal and transverse scans. (3) Resistivity measurements on samples from batches prepared in an identical manner all exhibit a single sharp transition at $T=0.4$~K. (4) The measured residual resistivity ratios are on the order of 20:1, and quantum oscillations have been observed in the thermopower \cite{Mun_2013} and magnetoresistance data \cite{Mun_thesis}. (5) Previous neutron powder diffraction measurements found no evidence of chemical disorder in identically grown YbBiPt samples. (6) We made measurements on two crystals of YbBiPt (selected from the batches used for our neutron scattering measurements) using high energy (232~keV) x-ray diffraction to probe the bulk of the crystals and found no second phase coherent with the YbBiPt chemical lattice.    Nevertheless, the apparent onset of short-range magnetic correlations at $T^*$ is surprising since it is well above $T_{\rm{N}}$ and, to the best of our knowledge, no distinct signature of this feature has been previously reported.  Given the relatively small ordered moment and the sizable broadening of the magnetic peaks, it is also now clear why previous neutron powder diffraction measurements failed to detect the magnetic order \cite{Robinson_1994}.

To check whether the scattering at $\bm\tau_{\rm{AFM}}$ arises from low-energy magnetic fluctuations rather than static order we performed constant-\textbf{Q} energy scans for  $T$ = 0.08~K on SPINS at \textbf{Q} = ($\frac{1}{2}~\frac{1}{2}~\frac{3}{2}$), and at positions well separated from the AFM Bragg position to capture the incoherent scattering background.  These data are shown in Fig.~\ref{Fig2}(d), where the shaded area corresponds to the additional magnetic scattering at the AFM position. The lines in Fig.~\ref{Fig2}(d) represent Gaussian fits with measured values for the FWHM of $\Delta E$ = 0.088(5), 0.090(7), and 0.089(6) meV, for \textbf{Q} = ($\frac{1}{2} \frac{1}{2} \frac{3}{2}$), (0.4~0.4~1.3), and (0.31~0.31~1.5), respectively. The instrumental energy resolution was determined from the FWHM of the elastic incoherent scattering from plastic to be $\Delta E$ = 0.087(1) meV and is indicated by the horizontal bar. We conclude that the peaks shown in Figs.~\ref{Fig2}(a) and (b) are elastic within our current experimental resolution, although we can not exclude that the scattering is quasielastic on an energy scale much smaller than 0.09 meV.  We note this possibility because all of the relevant energy scales in YbBiPt are on the order of the present energy resolution. However, the systematic absence of scattering at the ($\frac{h}{2} \frac{h}{2} \frac{h}{2}$) points would require that any quasielastic fluctuations be longitudinal (e.g. in the magnitude of the moment), and the absence of any change in the magnetic correlation lengths with temperature would be puzzling.

Evidence for unusual magnetic order in close proximity to a QCP has been found for other strongly-correlated materials fitting either the conventional SDW or Kondo-breakdown scenarios.  For example,   CeCu$_{6-x}$Au$_x$ exhibits dynamic short-range magnetic correlations for $x = x_{\rm{c}}$ =~0.1 (the critical concentration where non-Fermi-liquid behavior is clearly observed) \cite{Stockert_1998}.  For $x$ =~0.2, static short-range AFM order coexists with long-range AFM order at a different propagation vector \cite{Lohneysen_1998}, and persists well above $T_{\rm{N}}$ derived from specific-heat and AC-susceptibility measurements \cite{Stockert_1997}.  This is similar to the temperature dependence of the broad scattering component in YbBiPt. The presence of both broad and narrower components of the magnetic scattering may arise from a number of sources including the competition between magnetic and nonmagnetic ground states, the possible frustration inherent to the sublattice of side-sharing tetrahedra of Yb moments, as well as fluctuations associated with the nearby QCP.

\begin{acknowledgments}
We thank R. Flint for useful discussions and gratefully acknowledge the Missouri University Research Reactor,  Helmholtz-Zentrum Berlin f{\"{u}}r Materialien und Energie, the National Institute of Standards and Technology, and the U.S. Dept. of Commerce for the allocated beamtime and support during the experiments. Work at the Ames Laboratory was supported by the Department of Energy, Basic Energy Sciences, Division of Materials Sciences \& Engineering, under Contract No. DE-AC02-07CH11358.
\end{acknowledgments}

\end{document}